\documentclass[conference,9pt]{IEEEtran}
\IEEEoverridecommandlockouts
\usepackage{balance}
\usepackage{pifont}
\usepackage{stmaryrd}
\usepackage{optidef}
\usepackage{amssymb}
\usepackage{mathrsfs}
\usepackage{amsfonts}  
\usepackage[bookmarks=false]{hyperref}
\usepackage[dvips]{color}
\usepackage{xcolor}
\usepackage{epsfig,amssymb,amsmath,amsthm,bm}
\usepackage{graphicx}
\usepackage{algorithm}
\usepackage{cuted}

\usepackage{algorithm}
\usepackage{algpseudocode}

\usepackage{booktabs}

\usepackage{dsfont}
\usepackage{esint}
\usepackage{subfig}
\usepackage{subcaption}
\usepackage{url}
\usepackage{relsize}\usepackage{subdepth}\allowdisplaybreaks

\usepackage{cite}

\newcommand{\beq}{\begin{equation}}
\newcommand{\eeq}{\end{equation}}
\newcommand{\beqn}{\begin{eqnarray}}
\newcommand{\eeqn}{\end{eqnarray}}

\def\BibTeX{{\rm B\kern-.05em{\sc i\kern-.025em b}\kern-.08em
    T\kern-.1667em\lower.7ex\hbox{E}\kern-.125emX}}
\begin{document}

\title{Antenna Failure Resilience: Deep Learning-Enabled Robust DOA Estimation with Single Snapshot Sparse Arrays
\thanks{This work was supported in part by U.S. National Science Foundation (NSF) under Grants CCF-2153386 and ECCS-2340029.} 
}
\author{
	\IEEEauthorblockN{ Ruxin~Zheng$^{\dagger}$, Shunqiao~Sun$^{\dagger}$, Hongshan~Liu$^{\dagger}$, Honglei~Chen$^{\ddagger}$, Mojtaba~Soltanalian$^{\S}$, and Jian~Li$^\P$}\\
\IEEEauthorblockA{
$^{\dagger}$Department of Electrical and Computer Engineering, The University of Alabama, Tuscaloosa, AL 35487\\	
$^{\S}$Department of Electrical and Computer Engineering,  University of Illinois Chicago, Chicago, IL 60607 \\
$^{\P}$Department of Electrical and Computer Engineering, University of Florida, Gainesville, FL 32611\\ 
$^{\ddagger}$Mathworks, Natick, MA 01760
}
}

\maketitle

\vspace{0.5em}
\begin{abstract} 
Recent advancements in Deep Learning (DL) for Direction of Arrival (DOA) estimation have highlighted its superiority over traditional methods, offering faster inference, enhanced super-resolution, and robust performance in low Signal-to-Noise Ratio (SNR) environments. Despite these advancements, existing research predominantly focuses on multi-snapshot scenarios, a limitation in the context of automotive radar systems which demand high angular resolution and often rely on limited snapshots, sometimes as scarce as a single snapshot. Furthermore, the increasing interest in sparse arrays for automotive radar, owing to their cost-effectiveness and reduced antenna element coupling, presents additional challenges including susceptibility to random sensor failures. This paper introduces a pioneering DL framework featuring a sparse signal augmentation layer, meticulously crafted to bolster single snapshot DOA estimation across diverse sparse array setups and amidst antenna failures. To our best knowledge, this is the first work to tackle this issue. Our approach improves the adaptability of deep learning techniques to overcome the unique difficulties posed by sparse arrays with single snapshot. We conduct thorough evaluations of our network's performance using simulated and real-world data, showcasing the efficacy and real-world viability of our proposed solution. The code and real-world dataset employed in this study are available at \url{https://github.com/ruxinzh/Deep_RSA_DOA}. 
\end{abstract}

\smallskip
\section{Introduction}
Radar technology has become an essential component in the advancement of autonomous driving systems, particularly due to its robust performance in adverse weather conditions \cite{SUN_SPM_Feature_Article_2020, Markel_book_2022,Ruxin_TAES_2023}. Automotive radar systems, supporting the complex demands of autonomous vehicles, must provide high-resolution, four-dimensional (4D) data encompassing range, Doppler shifts, azimuth, and elevation angles, all while remaining cost-effective for mass production \cite{Sun_JSTSP_2021}. Although foundational aspects like the radar's range and Doppler resolution are determined by the waveform's bandwidth and the coherent processing interval respectively, a pivotal advancement lies in enhancing angular resolution for precise localization and tracking. MIMO (Multiple Input, Multiple Output) radar, which has become the industry standard for automotive applications, significantly contributes to this improvement. The angular resolution in MIMO radar is determined by the virtual array aperture size, which effectively enlarges the aperture beyond the physical dimensions of the receive antenna array. This capability can be further enhanced using super-resolution Direction of Arrival (DOA) estimation methods.

Confronting the obstacle of attaining substantial antenna aperture sizes for enhanced angular resolution, particularly in the context of filled arrays which require a significant number of antennas, sparse arrays have risen as an efficient and economical solution within the realm of automotive radar systems\cite{SUN_ICASSP_2020, Sun_JSTSP_2021, xu2023automotive,Zheng_RadarConf_2023}. Sparse arrays facilitate larger apertures and superior angular resolution with fewer elements and mitigate mutual coupling owing to their expansive element spacing, thereby offering a compelling alternative. Nonetheless, the design of optimal sparse arrays continues to be a formidable challenge, as the ideal configuration is intricately tied to specific, diverse requirements, indicating the absence of a one-size-fits-all solution for sparse array design\cite{Zheng2023asilomarSparse,lin2022design}. Furthermore, the occurrence of random sensor failures can lead to unpredictably sparse array geometries, complicating the scenario further. 

DOA estimation, a critical element in sensor array signal processing, finds extensive use across diverse fields including radar, sonar, navigation, and wireless communications, underscoring its universal applicability and importance \cite{Jian_07,Patole_SPM_2017,engels2017advances}. Despite extensive research and the development of numerous algorithms, most studies have traditionally focused on conditions with plentiful snapshots. This approach does not align well with the fast-paced and dynamic automotive environments, where the availability of radar sensor array snapshots is typically limited to a few or, in the most challenging situations, even a single snapshot.

Considering the snapshot limitations typical of automotive radar systems, traditional DOA estimation algorithms, reliant on accurate covariance matrix estimations, encounter notable challenges. This category encompasses parametric subspace-based methods such as the Multiple Signal Classification (MUSIC)\cite{schmidt1982signal} and the Estimation of Signal Parameters via Rotational Invariant Techniques (ESPRIT)\cite{Kailath_ESPRIT_1989}, along with beamforming techniques like the Minimum Power Distortionless Response (MPDR) beamformer and the Minimum Variance Distortionless Response (MVDR) beamformer, commonly known as the Capon beamformer\cite{capon1969high,van2002detection}. These methods depend heavily on an accurate estimation of the signal covariance matrix, which in turn requires a sufficient number of snapshots to achieve. Consequently, their effectiveness is considerably diminished in single-snapshot scenarios, which are prevalent in the dynamic conditions of automotive radar applications.

In the realm of single-snapshot super-resolution DOA estimation, Compressive Sensing (CS)\cite{donoho2006compressed} and IAA\cite{Yardibi_IAA_2010,Roberts_IAA_2010}, an iterative, nonparametric, and robust method, have emerged as notable methodologies. CS, exploiting the sparse representation of targets in the angular domain, and IAA, with its iterative, nonparametric method, both demonstrate exceptional enhancement capabilities. However, these techniques entail significant computational efforts which may restrict their utility in real-time applications due to the intensive processing involved.

Recently, deep learning (DL) strategies for DOA estimation have surged in popularity\cite{papageorgiou2021deep,fuchs2022machine,feintuch2023neural,gall2020spectrum,gall2020learning,9827881,eamaz2023automotive}, offering rapid inference, enhanced super-resolution, and efficacy in low signal-to-noise ratio (SNR) environments\cite{papageorgiou2021deep}. Despite the advantages, the predominantly data-driven nature of DL methods raises issues regarding their interpretability. In response, model-based deep learning approaches\cite{shlezinger2022model, shlezinger2023model, zheng2023interpretable, Zheng_EUSIPCO_2023,10052106} seek to merge the robustness of traditional mathematical models with the versatility of data-driven techniques, utilizing domain knowledge and mathematical frameworks to create interpretable, problem-specific solutions. Yet, the performance of such model-based techniques, when faced with unfamiliar sparse array configurations and an indeterminate number of sources, remains a challenge due to their reliance on deep learning principles.
Consequently, the quest for developing resilient, high-efficiency deep learning frameworks that can seamlessly adapt to a range of sparse array configurations, without necessitating retraining for each unique arrangement, is of paramount importance. Additionally, the capability of these models to accommodate random sensor failures\cite{8969105,4286013,8462643,4084805} is crucial for preserving the reliability and integrity of automotive radar systems.

In this paper, we introduce a novel deep learning framework for DOA estimation that features a sparse signal augmentation model with a unique augmentation layer, which randomly masks input signals to simulate various sparse array structures. This model is enhanced by incorporating domain-specific features such as sparse signal frequency embedding and active antenna position encoding, significantly advancing sparse array DOA estimation. Our comprehensive experiments with both simulated and real-world data demonstrate the framework's adaptability to different array configurations and its ability to handle the consequences of sensor failures, offering a robust and reliable solution for automotive radar systems. This approach not only improves generalizability and robustness but also addresses the unique challenges of sparse array DOA estimation, contributing a novel aspect to the field.

\section{System Model}\label{sigm}

Consider a scenario involving $K$ narrowband, far-field source signals, denoted as $s_k$ for $k = 1, \dots, K$, arriving at a linear, omnidirectional antenna array with $N$ elements from directions $\theta_k$. The temporal differences among the sensor outputs are represented by phase shifts, yielding the data model:
\begin{equation}
\begin{aligned}
\mathbf{y}(t) &= \sum_{k=1}^K \mathbf{a}(\theta_k)s_k(t) + \mathbf{n}(t)\\
&= \mathbf{A}(\theta) \mathbf{s}(t) + \mathbf{n}(t), \quad t = 1, \dots, T,
\end{aligned}
\end{equation}
where $t$ indexes the time snapshot, $\mathbf{n}$ denotes the complex $N \times 1$ white Gaussian noise vector, and $\mathbf{A}(\theta) = \left[\mathbf{a}(\theta_1), \mathbf{a}(\theta_2), \dots, \mathbf{a}(\theta_K)\right]$ represents the $N \times K$ array manifold matrix. Each element of $\mathbf{a}(\theta)$ is given by:
\begin{align}
\mathbf{a}(\theta) = \left[1, e^{\frac{2\pi d_2}{\lambda}\sin{\theta}}, \dots , e^{\frac{2\pi d_N}{\lambda}\sin{\theta}}\right]^T,
\end{align}
where $d_n$ specifies the spacing between the $n$-th element and the first element, and $\mathbf{s} = [s_1, s_2, \dots, s_K]^T$ is the vector of source signals. This paper focuses on estimating the directions of arrival (DOAs), $\theta$, using a single snapshot $\mathbf{y}$ of the array's response. Thus, with $T$ set to 1, the model simplifies to:
\begin{equation}
\mathbf{y} = \mathbf{A}(\theta)\mathbf{s} + \mathbf{n}.
\end{equation}

\begin{figure} 
\centering
\includegraphics[width= 0.4\textwidth]{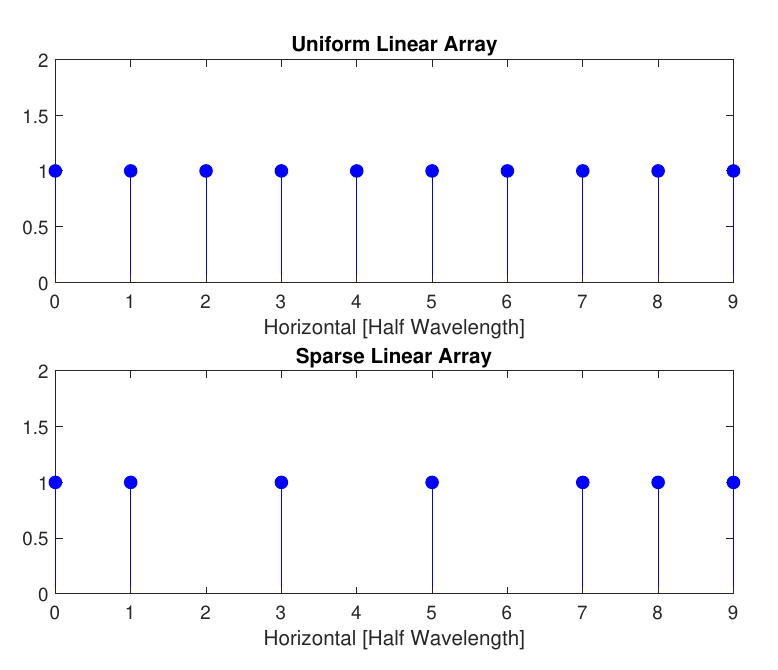}
\caption{\label{ULASLA} Example of ULA and SLA. The SLA has a 0.3 sparsity.
}
\end{figure}

Depending on performance and cost considerations, a Sparse Linear Array (SLA) can be employed for direction finding. Sparse arrays not only reduce hardware expenses but also diminish mutual coupling effects among antennas. This is because the spacing between elements in the receiver arrays is sufficiently large. Figure \ref{ULASLA} illustrates configurations of a 10-element Uniform Linear Array (ULA) and a 7-element SLA. Let $\lambda$ represent the wavelength of the carrier frequency. In the ULA, antennas are placed at grid points spaced at intervals of $9.5\lambda$, with each inter-element spacing being half a wavelength. The SLA, maintaining the same antenna aperture size as the ULA, can be conceptualized as a ULA modified by a binary mask. The sparsity of the SLA, defined as: 
\begin{equation}
\text{Sparsity} = 1 - \frac{N_{\text{SLA}}}{N_{\text{ULA}}},
\end{equation}
the ratio of the number of missing antenna elements in the SLA to those in the corresponding ULA, is 0.3 in this case.

\section{Deep-Learning Framework for DOA estimation}
\subsection{Network Architecture}

\begin{figure*} 
\centering
\includegraphics[width= 0.7\textwidth]{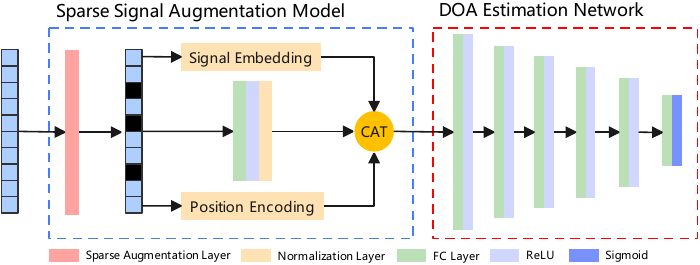}
\caption{\label{SADOA} Network design featuring sparse signal augmentation model coupled with DOA Estimation framework.
}
\vspace{-4mm}
\end{figure*}
\subsubsection{Sparse Augmentation Layer}
The technique of data augmentation\cite{shorten2019survey}, is employed in training deep learning networks to enhance model robustness and prevent overfitting. This is achieved by artificially expanding the dataset through various transformations. Common data augmentation techniques for computer vision tasks include flipping, rotation, and translation.

In the context of signal processing, the sparse augmentation layer is specifically designed to introduce controlled sparsity into the dataset. This layer generates a random binary mask that aligns with the size of the input signal. It includes a configurable parameter: the maximum allowed sparsity level, as detailed in section \ref{sigm}. This parameter governs the extent of sparsity by setting a cap on the number of elements in the input signal that can be zeroed. For example, consider a 10-element ULA. Setting the maximum sparsity to 0.3 allows the sparse augmentation layer to randomly zero out between zero to three elements of this array, thereby forming a sparse representation of the original signal. Additionally, this layer outputs the count of activated antenna elements, which is utilized for normalization. It is important to note that during the training phase, the number of activated antennas is determined by the sparse augmentation layer, while in the evaluation phase, it is decided through thresholding algorithms.

The sparsed signal is subsequently processed through a fully connected (FC) layer, followed by a ReLU activation layer. The function of the FC layer is mathematically defined as:
\begin{equation}
\text{output} = \mathbf{W} \times \text{input} + \mathbf{b},
\end{equation}
where \(\mathbf{W}\) represents the weight matrix, \(\text{input}\) is the incoming sparsed signal, and \(\mathbf{b}\) is the bias vector. Given the variability in sparsity across input signals, a normalization layer is essential to stabilize the output features. We define the function of the normalization layer as:
\begin{equation}
\text{output} = \frac{\text{input}}{N_{\text{SLA}}},
\end{equation}
where \( N_{SLA} \) is the count of non-zero elements in the sparse signal.

This normalization approach ensures that the output features are adjusted relative to the number of active inputs, thereby accommodating the inconsistent sparsity of the input signals and enhancing the model's reliability in feature representation.
 
\subsubsection{Domain Knowledge Crafted Features}
\begin{figure} 
\centering
\includegraphics[width= 0.48\textwidth]{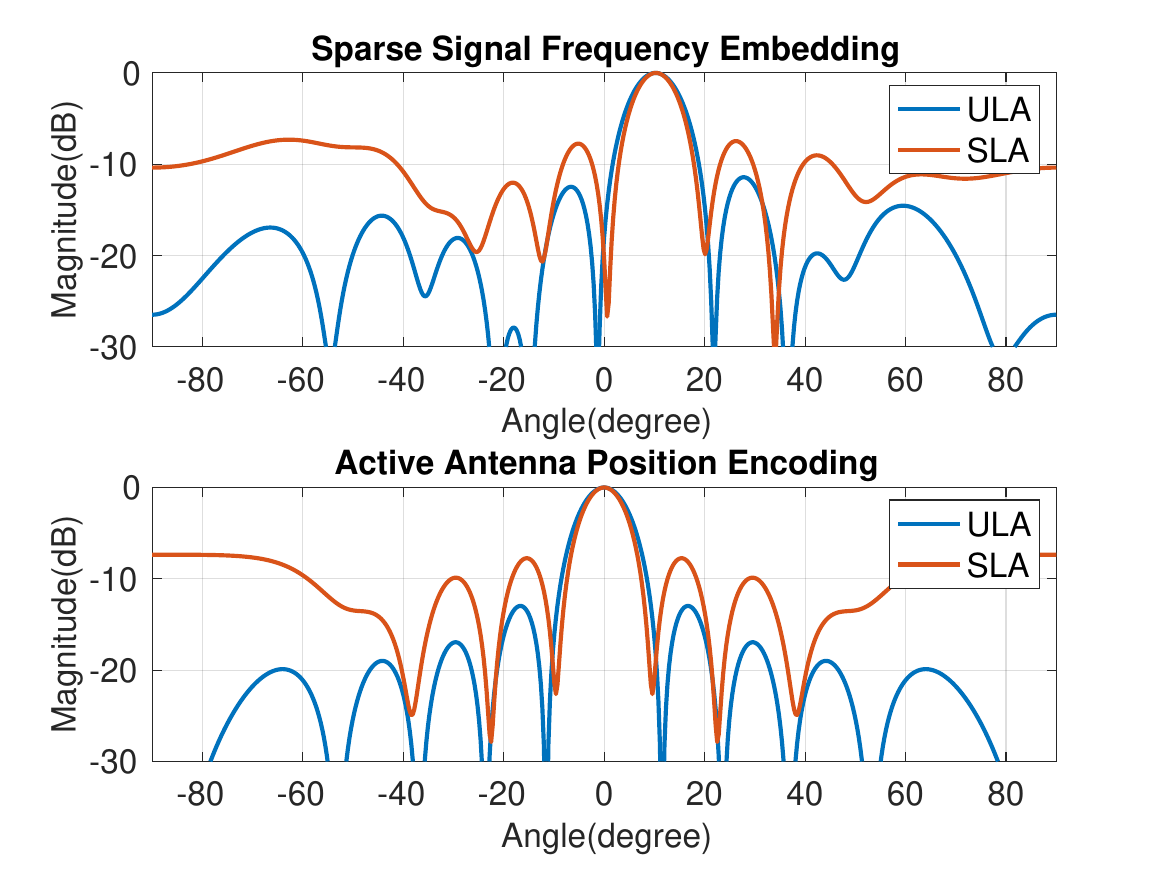}
\caption{\label{feature} Example of crafted features of ULA and SLA.  
}
\vspace{-4mm}
\end{figure}

Incorporating domain knowledge through handcrafted features is pivotal in the training of deep learning networks, significantly enhancing model performance by directly injecting expert insights and established heuristics, especially beneficial in complex or poorly understood domains. In this study, we employ two specifically crafted features: \textit{Sparse Signal Frequency Embedding} and \textit{Active Antenna Position Encoding}. These features involve transforming the sparse signal and the position of active antenna elements into the frequency domain, respectively. The embedding process is mathematically defined as:
\begin{equation}
\text{output} = \frac{\mathbf{A}^H \times \text{input}}{N_{\rm SLA}},
\end{equation}
where $\mathbf{A}^H$ represents the Hermitian transpose of the array manifold matrix, and $N_{SLA}$ denotes the number of activate antenna. Figure \ref{feature} provides an illustrative example of these embeddings for a 10 dB SNR signal targeting a single object at a 10-degree angle, utilizing both ULA and SLA configurations. The array setups are depicted in Figure \ref{ULASLA}. Different array geometries yield distinct features, evidenced by the variations in peak side lobe level and main lobe beamwidth across the frequency domain spectra.

\subsubsection{DOA Estimation Network}\label{doa}
DOA estimation can be achieved using various types of deep learning networks. In this paper, we specifically focus on developing a framework for sparse array DOA estimation and therefore choose an MLP due to its simplicity. However, this DOA estimation network could be replaced with other types of networks such as RNNs (Recurrent Neural Networks) or CNNs (Convolutional Neural Networks). The DOA estimation network we employ consists of six FC layers. The output sizes for the first five FC layers are 2048, 1024, 512, 256, and 128, respectively, each followed by a ReLU activation layer. The final FC layer, which serves as our output layer, has an output size determined by the desired angle scanning grid size and is followed by a Sigmoid activation layer.

\subsection{Data Generation and Labeling}
We utilize a ULA consisting of $10$ elements with inter-element spacing of half-wavelength to simulate signals for maximal $3$ targets with a minimum separation of $\Delta \phi = 1^{\circ}$. The radar field of view (FOV) is set as $\boldsymbol{\phi}_{\rm FOV} = [-30^{\circ},30^{\circ}]$, which is discretized with a step size of $1^{\circ}$, resulting in a grid $G \in \mathbb{R}^{1 \times M}$ with $M = 61$ possible DOA angles. Reflection coefficients $s$ for each DOA source are generated as random complex numbers. Signals are labeled according to

\begin{equation}\label{bvgt}
  {\bf GT}_{n} =
    \begin{cases}
      |s_k|, & \text{if } \theta_k = G_n\\
      0, & \text{else}
    \end{cases} 
    \;\;{\rm for \;\;} n = 1, 2, \cdots, M.
\end{equation}

For simulation, we randomly select a number of targets ranging from 1 to 3 and generate $100,000$ at various SNR levels from $0$ dB to $30$ dB, in $5$ dB increments, for the training dataset. For the validation set, we employ the same configuration and generate $1,000$ signals for each SNR levels. 

\subsection{Training Approach}
The proposed network underwent training over \(200\) epochs with a batch size of \(1024\), utilizing the Adam optimizer at a learning rate of \(0.0001\) and employing a Binary Cross-Entropy (BCE) loss function. The model was trained end-to-end, which involved direct training from input to output without any intermediate pre-processing or post-processing steps. The primary objective of this training regimen was to minimize the BCE loss and enhance the accuracy of predictions. This experiment was conducted using Google Colab. To mitigate overfitting, a validation set was employed, and the model configuration yielding the lowest validation loss was selected for subsequent performance evaluations, as detailed in Section \ref{perf}.
 
\subsection{Real World Dataset}
Currently, there is no publicly available real-world dataset for DOA estimation; existing models are trained and evaluated using simulated datasets. To address this gap, we developed a DOA estimation dataset in a parking lot scenario. A stationary vehicle equipped with a TI cascade imaging radar \cite{TI_Cascade} collected data from a corner reflector positioned 15 meters away, capturing signals from all possible directions. This process generated 195 high-SNR signals from various angles, each representing a single target. By superimposing these vectors, we simulated scenarios with multiple targets. Notably, this real-world dataset was not used for training purposes but solely to demonstrate our network's performance during testing.

\section{Performance Evaluation}\label{perf}
We evaluate our proposed model based on three critical aspects of DOA estimation: accuracy, separability, and complexity. The evaluations for accuracy and separability are conducted under two scenarios: using ULA and SLA, with the latter's sparsity set at 0.3, and randomly generated in each Monte Carlo trial. To ensure a thorough comparison, our model's performance is benchmarked against traditional DOA estimation methods such as IAA and digital beamforming (DBF) implemented via Fast Fourier Transform (FFT). Additionally, a MLP sharing the same network structure as described in Section \ref{doa}, but with adjustments for different input sizes, is also compared. The scanning angle grid for IAA and DBF, ranging from $-30^{\circ}$ to $30^{\circ}$, is discretized into a 61-point grid to match the output resolution of the deep learning networks. The maximum number of iterations for IAA is capped at 15, beyond which performance gains are minimal \cite{Yardibi_IAA_2010}. All tests are conducted over 5,000 Monte Carlo trials.

\subsection{Accuracy}
We adopt the Mean Squared Error (MSE) as the performance metric to evaluate the accuracy of DOA estimation methods. Our approach utilizes a conventional grid-based method where the DOA estimates are derived from the estimated spectrum via peak search. The grid-induced error, depicted by the dark dashed line in the accompanying charts, is quantified by the MSE between the source DOA and the nearest grid angle. This error represents a fundamental lower bound for this metric.

\subsubsection{Single Target}
In each Monte Carlo trial, a single off-grid source is simulated with a direction randomly chosen from the interval \([-30^\circ, 30^\circ]\), accompanied by its corresponding SNR.
\begin{figure} [h]
\centering
\includegraphics[width=0.48\textwidth]{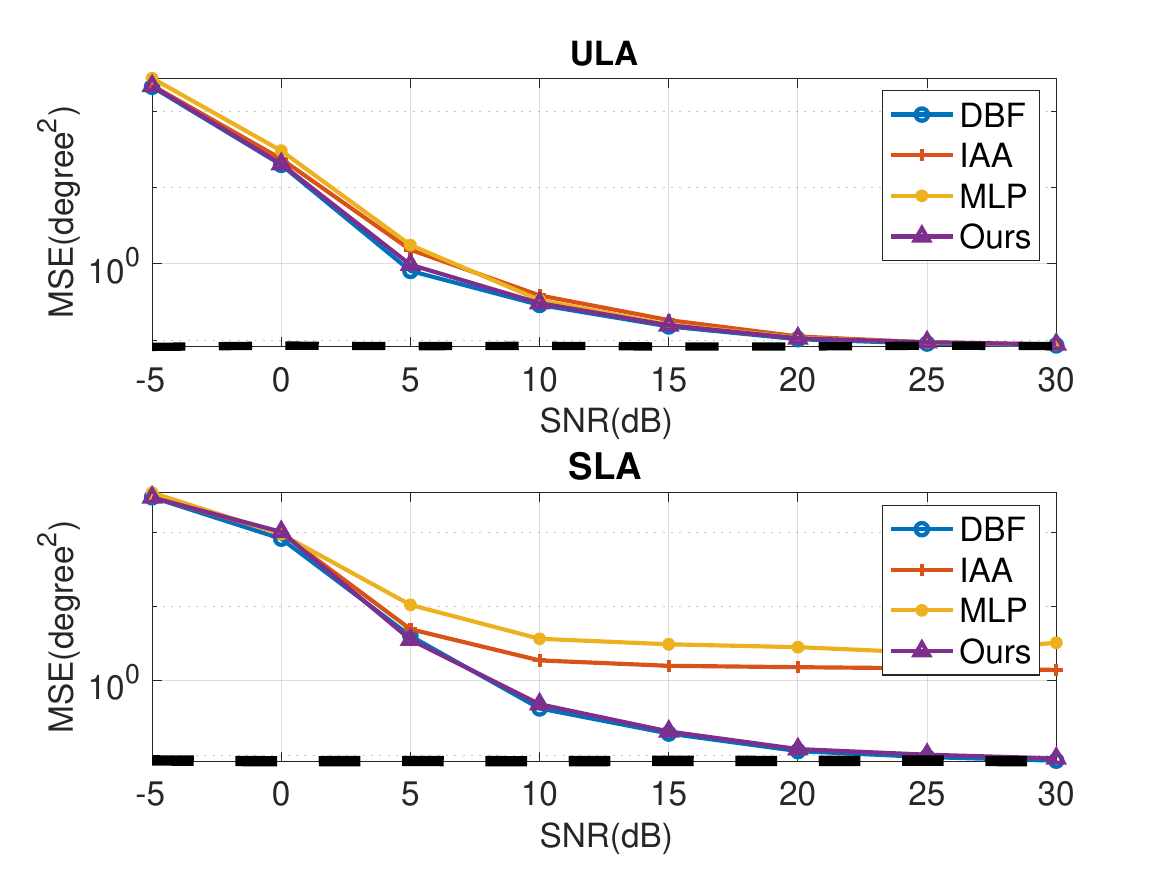}
\caption{Logarithmic scale MSE versus SNR in the DOA estimation of a single, randomly generated off-grid target.}
\label{acc1}
\vspace{-4mm}
\end{figure}
As depicted in Figure \ref{acc1}, all DOA estimation methods exhibit similar performance for the ULA configuration. However, for the SLA, the IAA and the MLP show significant performance degradation, highlighting their sensitivity to data sparsity. Conversely, DBF and our proposed network demonstrate robustness against SLA-induced sparsity.

\subsubsection{Two Targets}
Each Monte Carlo trial involves simulating two off-grid sources with directions randomly drawn from the intervals \([-0.6^\circ, 0.4^\circ]\) and \([9.6^\circ, 10.4^\circ]\), each with its respective SNR.
\begin{figure} [h]
\centering
\vspace{-0mm}
\includegraphics[width=0.48\textwidth]{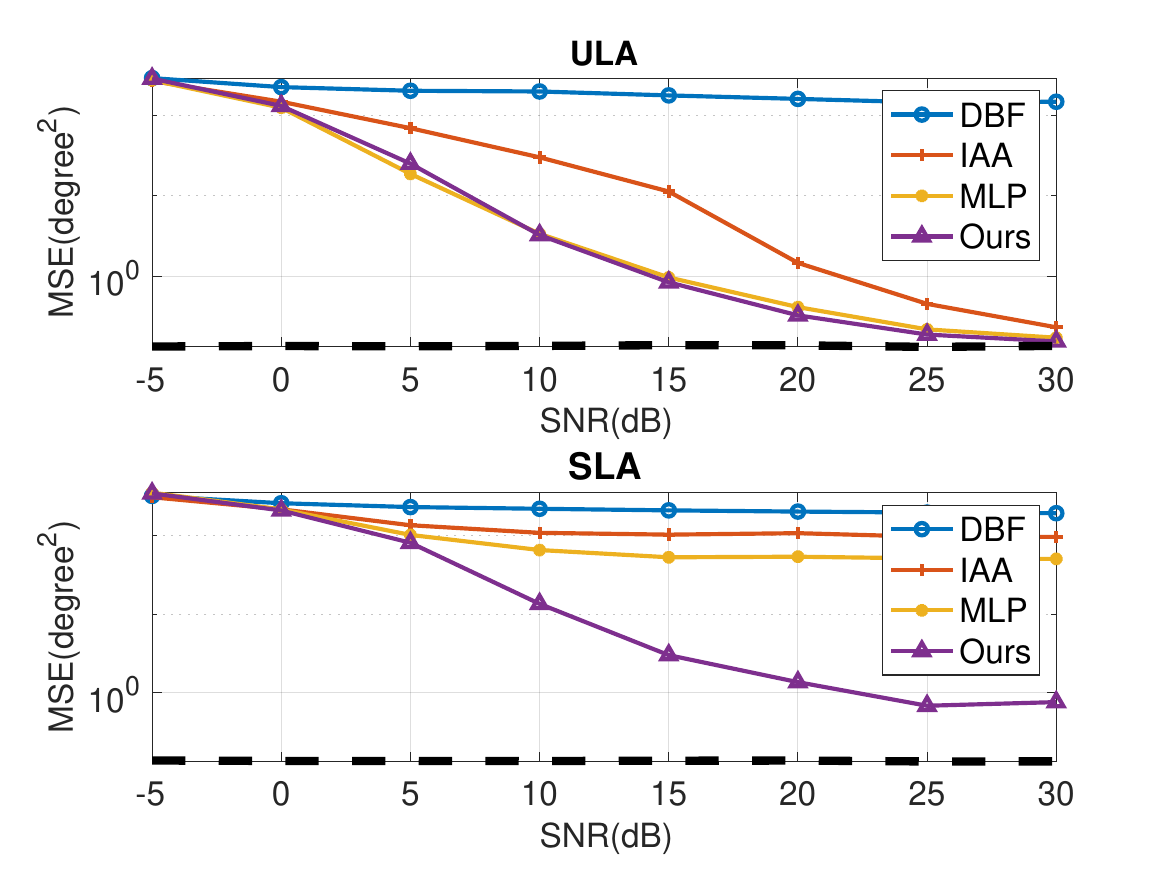}
\caption{Logarithmic scale MSE versus SNR for DOA estimation of two randomly generated off-grid targets, with the first target drawn from the interval \([-0.6^\circ, 0.4^\circ]\) and the second from \([9.6^\circ, 10.4^\circ]\).}
\label{acc2}
\vspace{-4mm}
\end{figure}

As shown in Figure \ref{acc2}, for the ULA configuration, DBF exhibits high MSE, struggling to resolve two closely spaced targets due to its limited resolution capability. The deep learning approaches outperform IAA, underscoring the potential of deep neural networks in DOA estimation. For the SLA, while DBF continues to show high MSE, IAA and MLP suffer substantially from the array's missing elements. Our proposed method not only performs optimally in the SLA but also demonstrates superior robustness under these challenging conditions.

\setcounter{figure}{6}
\begin{figure*}
\centering
\includegraphics[width= 0.99\textwidth]{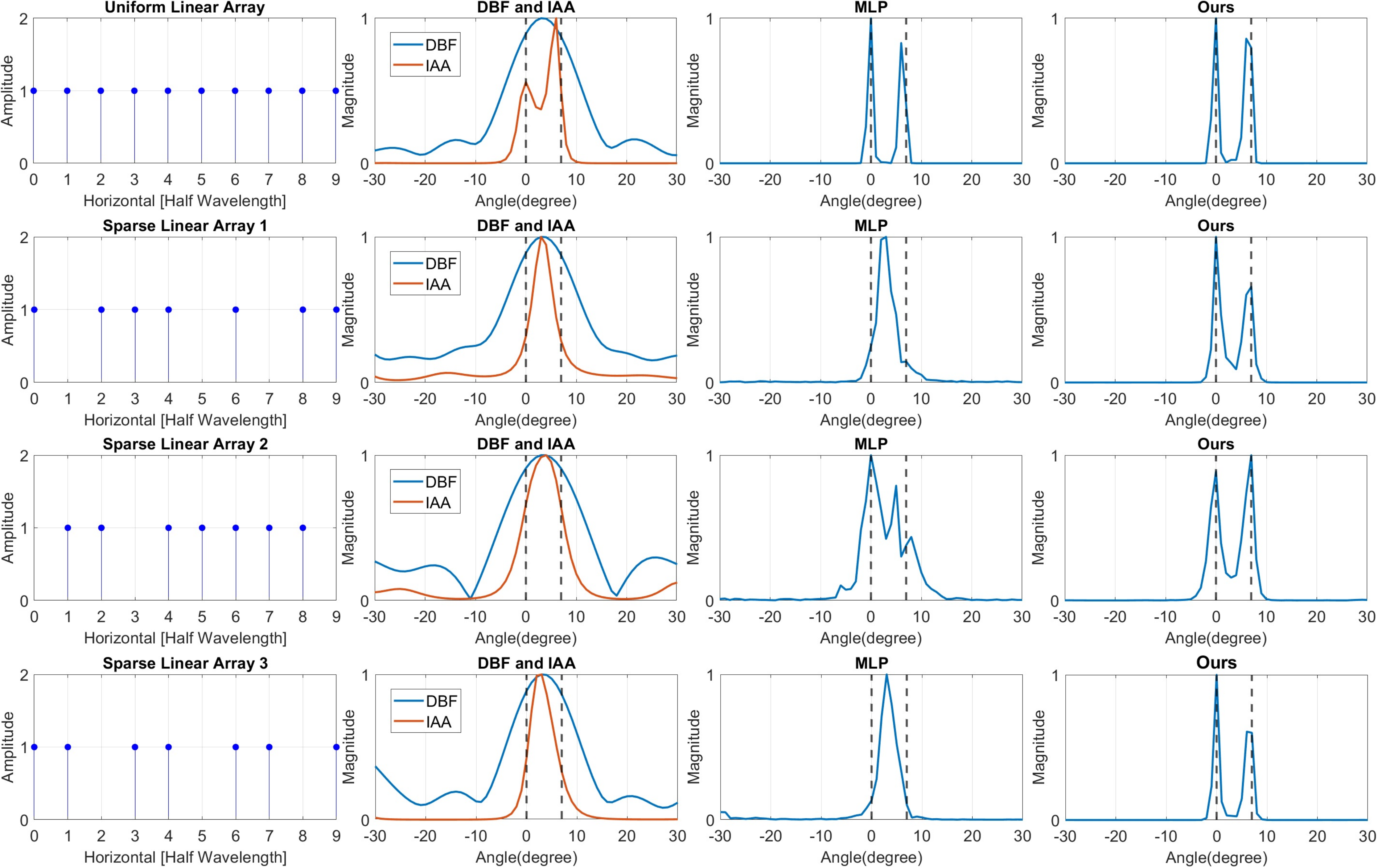}
\caption{Real world data examples.}
\label{example} 
\vspace{-5mm}
\end{figure*}

\subsection{Separability}
To assess the ability of our DOA estimation methods to resolve closely situated targets, we conducted an experiment featuring two targets positioned symmetrically around the origin, at angles \(-\Delta \theta / 2\) and \(+\Delta \theta / 2\), respectively. Here, \(\Delta \theta\) denotes the angular separation between the targets. A trial is classified as a ``hit'' if the deviation between the estimated DOAs and the actual positions is within \(\pm 1^\circ\). We compute the hit rate as the proportion of hits over 5,000 Monte Carlo trials, each at an SNR of 40 dB.

\setcounter{figure}{5}
\begin{figure}[h]
\vspace{-0mm}
\centering
\includegraphics[width=0.48\textwidth]{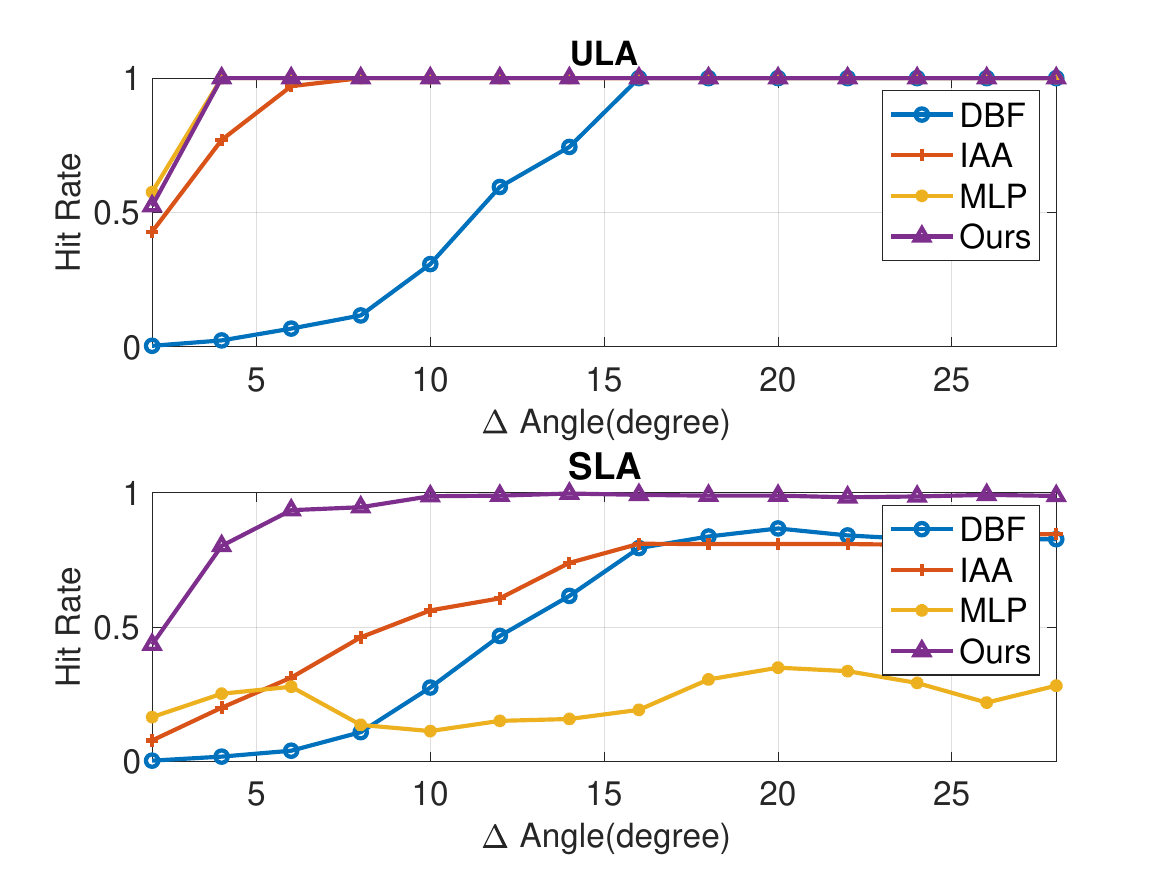}
\vspace{-2mm}
\caption{Hit rate as a function of angular separation \(\Delta \theta\).}
\label{sep}
\vspace{-4mm}
\end{figure}

Figure \ref{sep} illustrates the performance results. In the ULA configuration, the deep-learning methods significantly outperform the model-based algorithms in terms of separability, with DBF exhibiting the poorest performance. For the SLA, both MLP and IAA experience a notable decrease in hit rate, while our proposed method maintains superior separability compared to all other evaluated techniques.

\vspace{-1mm}
\subsection{Complexity Evaluation}
We conducted a comprehensive evaluation of the complexity of our proposed method by analyzing its inference time and the number of trainable parameters. For consistency and fairness in comparison, all DOA estimation methods were implemented on Google Compute Engine. The inference time for each method was averaged over $5,000$ trials. For the deep learning-based approaches, we utilized a batch size of one. As detailed in Table \ref{table_1}, the DBF method exhibited the shortest inference time, while both deep learning approaches, MLP and our proposed network, were more than ten times faster than the IAA. Our method utilizes 1.3 million more trainable parameters than MLP and requires only an additional 0.8 milliseconds of inference time per trial relative to MLP. These results suggest that our approach enhances robustness and improves performance in SLA without significantly increasing complexity or sacrificing performance in ULA.

\begin{table}[ht]
\centering
\resizebox{0.95\linewidth}{!}{%
\begin{tabular}{|l|c|c|} \hline 
\textbf{Methods} & \textbf{Inference Time (ms)} & \textbf{\# Trainable Parameters} \\ \hline

DBF  & $0.3$ & --\\

IAA  & $32.8$  & -- \\

MLP  & $2.3$ & $2,848,829 $  \\

Ours  & $3.1$ & $4,106,301$ \\\hline
\end{tabular}}
\vspace{0 mm}
\caption{Inference time comparison of DOA methods} 
\label{table_1}
\end{table}
\vspace{-3 mm}
\subsection{Qualitative Analysis}
To better demonstrate the efficacy of our proposed networks, we employed real-world data by superimposing two signals: one containing a target at \(0^\circ\) and another at \(7^\circ\), thus creating a composite signal with two distinct targets. The true DOAs are marked on Figure \ref{example} using black dashed lines. The first row presents the spectral outputs for a ULA configuration, where the DBF method fails to differentiate the two targets, whereas all other methods successfully resolve them.

The second, third, and fourth rows display the results for various SLA configurations. In these setups, DBF consistently fails to resolve the two targets, a limitation also observed with the IAA and the MLP across all SLA configurations. In contrast, our proposed method effectively resolves the targets in all SLA configurations, demonstrating its robustness and superior performance across diverse sparse array geometries. This underlines the significant capabilities of our network in handling complex signal environments.
\section{Conclusion}
This paper has introduced a novel DL framework designed to advance the field of DOA estimation, specifically tailored for automotive radar systems which often operate under the constraints of single snapshot scenarios and sparse array configurations. Our proposed method incorporates a unique sparse signal augmentation model, enabling robust DOA estimation under challenging conditions such as single snapshot, antenna failure, various sparse array geometries. Comprehensive evaluations using simulated data, along with qualitative analyses of real-world data, confirm that our approach consistently outperforms traditional methods. It delivers faster inference times, enhanced super-resolution capabilities, and robust performance in low SNR environments. These attributes make our framework particularly well-suited to the dynamic and demanding requirements of automotive radar systems, which necessitate both high angular resolution and exceptional reliability.

\bibliographystyle{IEEEtran}
 \balance
\bibliography{refs}

\end{document}